\documentclass[%
superscriptaddress,
 amsmath,amssymb,
 aps,
prb,
]{revtex4-1}

\usepackage {graphicx,epsfig,graphics,color}
\usepackage{dcolumn}
\usepackage{bm}
\usepackage{hyperref}
\usepackage{sidecap}
\usepackage{float}

\RequirePackage[normalem]{ulem} 
\RequirePackage{color}\definecolor{RED}{rgb}{1,0,0}\definecolor{BLUE}{rgb}{0,0,1}

\begin{document}
\title{Effect of topology on quasi-particle interactions in the Weyl semimetal WP$_2$}

\author{Dirk Wulferding}
\affiliation{Institute for Condensed Matter Physics, TU Braunschweig, Mendelssohnstr.~3, 38106 Braunschweig, Germany}
\affiliation{Laboratory for Emerging Nanometrology (LENA), TU Braunschweig, Langer Kamp~6, 38106 Braunschweig, Germany}

\author{Peter Lemmens}
\affiliation{Institute for Condensed Matter Physics, TU Braunschweig, Mendelssohnstr.~3, 38106 Braunschweig, Germany}
\affiliation{Laboratory for Emerging Nanometrology (LENA), TU Braunschweig, Langer Kamp~6, 38106 Braunschweig, Germany}

\author{Florian B{\"u}scher}
\affiliation{Institute for Condensed Matter Physics, TU Braunschweig, Mendelssohnstr.~3, 38106 Braunschweig, Germany}
\affiliation{Laboratory for Emerging Nanometrology (LENA), TU Braunschweig, Langer Kamp~6, 38106 Braunschweig, Germany}

\author{David Schmeltzer}
\affiliation{Physics Department, City College of the City University of New York, New York 10031, USA}

\author{Claudia Felser}
\affiliation{Max Planck Institute for Chemical Physics of Solids, N\"{o}thnitzer Str.~40, 01187 Dresden, Germany}

\author{Chandra Shekhar}
\affiliation{Max Planck Institute for Chemical Physics of Solids, N\"{o}thnitzer Str.~40, 01187 Dresden, Germany}

\date{\today}

\begin{abstract}
We compare two crystallographic phases of the low-dimensional WP$_2$ to better understand features of electron-electron and electron-phonon interactions in topological systems. The topological $\beta$-phase, a Weyl semimetal with a giant magneto-resistance, shows a larger intensity of electronic Raman scattering compared to the topologically trivial $\alpha$-phase. This intensity sharply drops for $T < T^* = 20$ K which evidences a crossover in the topological phase from marginal quasiparticles to a coherent low temperature regime. In contrast, the non-topological $\alpha$-phase shows more pronounced signatures of electron-phonon interaction. Here there exist generally enlarged phonon linewidths and deviations from conventional anharmonicity in an intermediate temperature regime. These effects provide evidence for an interesting interplay of electronic correlations and electron-phonon coupling. Both interband and intraband electronic fluctuations are involved in these effects. Their dependence on symmetry as well as momentum conservation are critical ingredients to understand this interplay.


\end{abstract}

\maketitle

\section{Introduction}

Recently, the aspect of topology in condensed matter physics has gained immense importance, leading to the experimental discovery of novel phases -- topological insulators, Dirac semimetals, and Weyl semimetals -- in systems with strong spin-orbit coupling~\cite{manna-18, yan-17}. These emergent topological phases are governed by Chern numbers and symmetry aspects of the electronic band structure close to the Fermi energy. For example, Weyl semimetals require broken time-reversal or inversion symmetry and non-centrosymmetric symmetry with both symmetries broken. Otherwise, they can be classified either as Dirac semimetals with relativistic band states or as topologically trivial.

While so far the focus has been mainly on single particle physics, the importance of interactions, such as electronic correlations or electron-phonon interaction, has been emphasized recently. The former are relevant, e.g., in certain pyrochlores~\cite{kim-12} while the latter could be relevant if there is a matching of typical phonon and electronic energy scales. Electron-phonon interactions can be used to characterize the electronic band topology or even induce a topological phase~\cite{garate-13,saha-13}. Related phonon anomalies have indeed been observed in Raman scattering (RS) of the Dirac semimetal Cd$_3$As$_2$~\cite{sharafeev-17} and the Weyl semimetal NbIrTe$_4$~\cite{Chen-2018}.


Here, we present a comparative Raman spectroscopic study of two structural modifications of WP$_2$, the topologically trivial $\alpha$- and the topologically relevant $\beta$-phase. The latter is a robust type-II Weyl semimetal, which means that Weyl points with tilted 3D cones exist at the intersection of hole and electron pockets. A pre-requisite for the existence of such Weyl points is a broken inversion symmetry~\cite{ruhl}. Furthermore, neighboring Weyl points in $\beta$-WP$_2$ have the same chirality which enhances their stability and robustness against lattice distortions. Indeed, no appreciable temperature dependence of the electronic band structure and its topological features has been observed in the temperature range 25 - 170~K~\cite{Razzoli-2018}. 

The $\beta$-phase evidences strong electron-electron correlations with a hydrodynamic regime of charge transport for $T$~$<$~$T^*$=20~K~\cite{gooth-18}. This transport is also highly anisotropic based on the chain-like arrangement of tungsten-phosphorus polyhedra. There exist a giant magnetoresistance of $4 \cdot 10^6$\%, a residual low-temperature resistivity of 3 n$\Omega$cm, and a charge carrier mean free path of 0.5 mm~\cite{kumar-17}. Band structure calculations of the monoclinic $\alpha$-phase point to a trivial electronic band structure with a center of inversion. Nevertheless, it shows an appreciably magnetoresistance and of $5 \cdot 10^5$\% and a residual low-temperature resistivity of 42 n$\Omega$cm~\cite{du-18}. These dissimilar properties with respect to correlation effects call for a detailed comparison of their relevant scattering mechanisms.


Our study shows quasielastic and finite energy electronic Raman scattering (ERS) in both phases which is usually a fingerprint of electronic correlations and/or nesting. In the topological $\beta$-phase the scattering at finite energies shows a drop in intensity for temperatures below approximately 50 K, that seems to initiate the electronic crossover to hydrodynamic charge transport at $T^* = 20$ K. This phase also shows a larger intensity of quasielastic scattering with a more pronounced $T$ dependence. We propose an important role of the highly anisotropic Fermi surfaces and that intraband and interband fluctuations evolve differently. Microscopic modeling of  Raman scattering (RS) supports these findings. Furthermore, phonons show decisive anomalies in frequency and linewidth with generally larger effects in the non-topological $\alpha$-phase. This demonstrates a remarkable interplay and competition of electron-phonon and electron-electron interactions in these phases.

\section{Experimental Details}

Single crystals of both phases of WP$_2$ were grown via chemical vapor transport~\cite{kumar-17}. The resulting crystal structure ($\alpha$- or $\beta$-phase) of the samples depends on a detailed temperature control during the growth process. Orthorhombic $\beta$-WP$_2$ crystallizes in the space group $Cmc21$, with one W- and two P-ions located on $4a$ Wyckoff positions and chains of tungsten polyhedra running along the crystallographic $a$ axis. Monoclinic $\alpha$-WP$_2$ crystallizes in the space group $C12/m1$, with one W- and two P-ions each located on $4i$ Wyckoff positions and chains of tungsten polyhedra running along the crystallographic $b$ axis~\cite{ruhl}. The reduced coordination of the latter structural element is also the basis of a quasi-1D character of the electronic states in both phases. Details of the observed phonon modes of both phases can be found in the Supplement. Our room temperature data are in excellent agreement with a previous study of phonons in $\beta$-WP$_2$~\cite{Su-2019}.

Temperature-dependent RS experiments probing excitations down to 100 cm$^{-1}$ have been performed using a Horiba LabRam HR800 spectrometer equipped with a holographic notch filter and an excitation wavelength $\lambda = 532$ nm. Further experiments with a lower cut-off energy of about 30 cm$^{-1}$ were carried out using a Dilor-XY triple spectrometer with an excitation wavelength of 488 nm.

\section{Results}

\begin{figure*}
\label{rt-spectra}
\centering
\includegraphics[width=16cm]{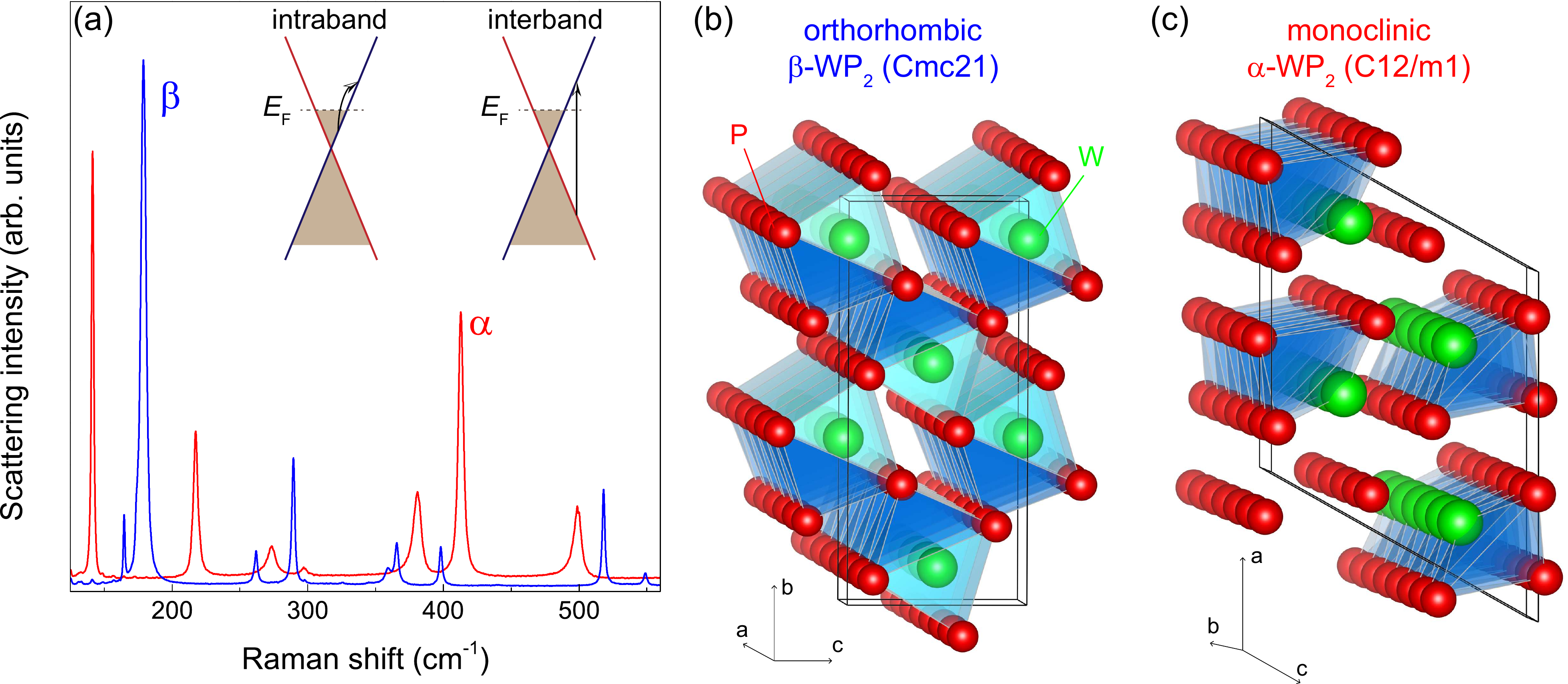}
\caption{(Color online) (a) Unpolarized room-temperature Raman spectra of the orthorhombic $\beta$-WP$_2$ (blue) and the monoclinic $\alpha$-WP$_2$ (red). The inset sketches intraband and interband scattering processes. (b) and (c) illustrate the crystal structure of both phases with strands of polyhedra forming along the crystallographic $a$ axis and the $b$ axis, respectively.}
\end{figure*}


\subsection{Phonon Scattering}

In Fig. 1(a) we compare Raman spectra of $\alpha$- and $\beta$-WP$_2$. Despite similarities, both phases have their very own distribution of phonon frequencies. This also highlights the phase purity of both crystals. We observe 7 out of 9 expected modes from $\alpha$-WP$_2$, and 13 out of 15 modes from $\beta$-WP$_2$. The missing modes are rationalized by their too small intensity or overlap with the observed ones. Please notice that the intensity of phonon scattering of the two phases is rather similar. This is consistent with a similar electronic polarizability on the energy scale of the incident photons, i.e. approximately 2~eV. Throughout the manuscript the data are not scaled to allow a better comparison of the two phases.

We notice a generally larger phonon linewidth in $\beta$-WP$_2$ compared to $\alpha$-WP$_2$. The mean linewidths are $\Delta\omega_{\mathrm{mean}}$=2.6~cm$^{-1}$ and 3.4~cm$^{-1}$, respectively (see also a list of low temperature data in Table I and II in the Supplement). It is tempting to attribute the larger linewidth to an enhanced electron-phonon interaction and respective scattering processes that limit the phonon lifetime in $\alpha$-WP$_2$. More details will be discussed further below.

In Figs. 1(b) and 1(c) sketches of the crystal structure of both phases are given. The ions are coordinated in chains of polyhedra. This structural element is also the basis of a quasi-1D character of the electronic states of both phases. The chains run along the crystallographic $a$ axis or along the $b$ axis for orthorhombic $\beta$ and monoclinic $\alpha$ WP$_2$, respectively.

\begin{figure*}
\label{phonons}
\centering
\includegraphics[width=16cm]{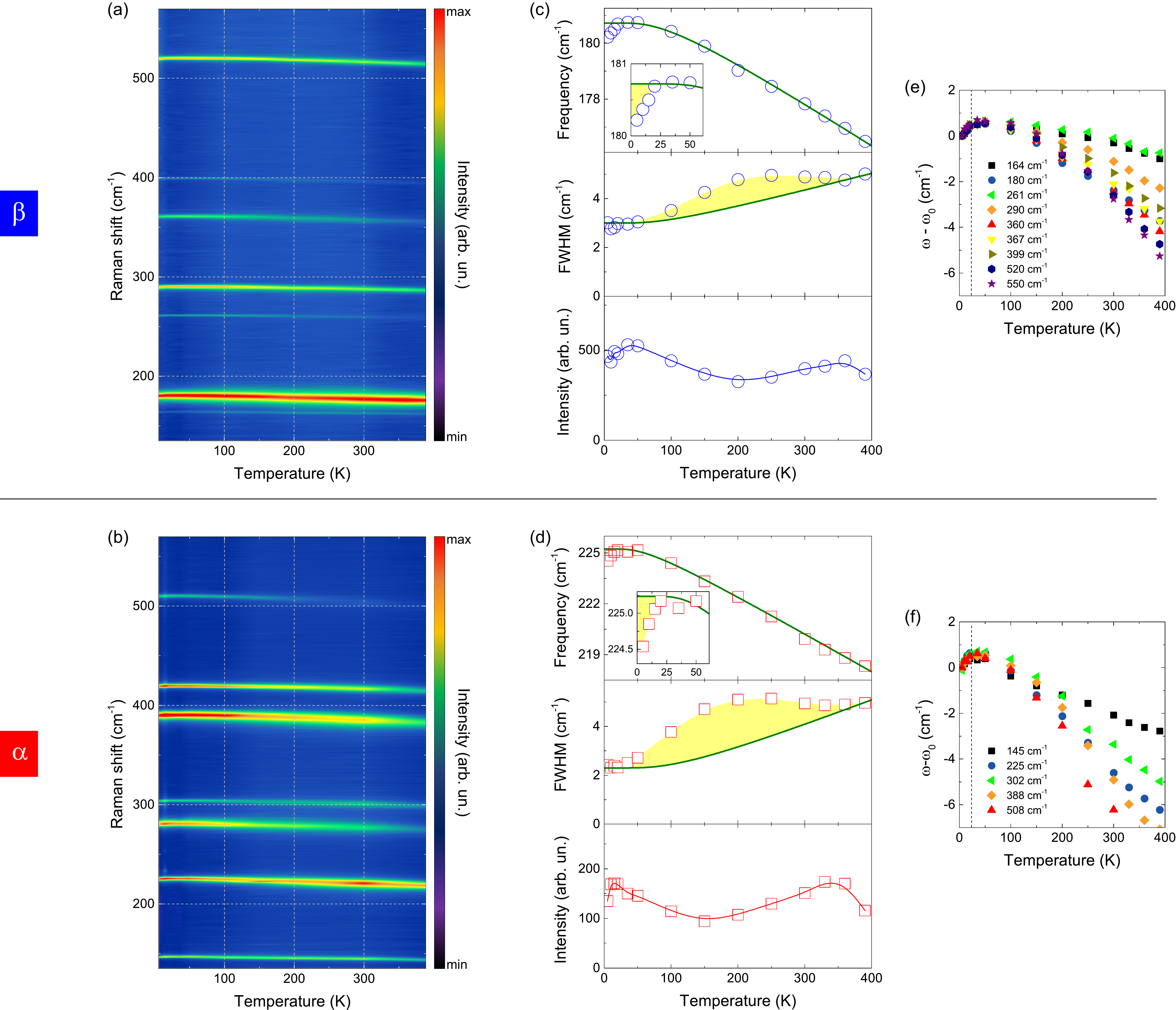}
\caption{(Color online) (a)-(b) Temperature dependent Raman spectra from the topological $\beta$-WP$_2$ and the non-topological $\alpha$-WP$_2$, respectively, as false color plots. (c)-(d) Detailed analysis of the phonon parameters of the 180 cm$^{-1}$-mode in $\beta$- and the 225 cm$^{-1}$-mode in $\alpha$-phase, respectively. (e)-(f) Normalized temperature dependence of phonon frequencies in $\beta$- and $\alpha$-WP$_2$, respectively.}
\end{figure*}

The temperature dependent Raman data plotted in Fig. 2(a) and 2(b) show a smooth and gradual evolution of the phonons, without any evidence for a structural phase transition. Therefore, $\alpha$-WP$_2$ retains its center of inversion, while $\beta$-WP$_2$ remains without. For a more detailed analysis, we plot the parameters frequency, linewidth, and intensity, of one selected phonon mode in Fig. 2(c) and (d). Both frequency ($\omega$) and linewidth ($\Gamma$) can be approximated by first-order anharmonicity~\cite{balkanski-83}, with $\omega (T) = \omega_0 - A \cdot \left( 1+\frac{2}{\mathrm{exp}(\hbar \omega_0 / 2k_B T) -1} \right)$, and $\Gamma (T) = \Gamma_0 \cdot \left( \frac{1+B}{\mathrm{exp}(\hbar \omega_0 / k_B T) -1} \right)$, where $A$ and $B$ are fitting parameters. This modeling corresponds to a decay of optical phonons into several acoustic phonons with pairwise compensating, large momenta.

While these functions result in an overall reasonable description of the data, there are two temperature regimes with deviations. At low temperatures, $T \lessapprox 20$ K, the phonon frequencies show an anomalous softening for both phases, see the insets in Fig. 2(c) and (d) for a zoom into the data. It could be tempting to assign this effect to the crossover to hydrodynamic charge transport in $\beta$-WP$_2$~\cite{gooth-18}. However, for $\alpha$-WP$_2$ with similar softening anomalous transport properties have not being reported. Therefore, a more general approach is appropriate. As both phases are low dimensional semimetals, enhanced electron-phonon coupling due to intraband fluctuations~\cite{du-18} and nesting are more plausible. The latter is related to the low dimensional Fermi surface, as we will discuss below.

The second anomalous regime is given at intermediate temperatures, 50 K $< T <$ 300 K. Here, the phonon linewidth shows an anomalous broadening, see the yellow-shaded area in Fig. 2(c) and (d), and the phonon intensities deviate from the expected monotonous behavior. In this regime of finite temperatures the anomalies may be attributed to a modification of anharmonicity, i.e. optical phonon decay processes into acoustic, zone boundary phonons. This includes all finite momentum self-energy effects that are relevant for the matrix elements of electron-phonon interaction. It is remarkable that the phonon anomalies, despite being similar for the two phases, are larger for the non-topological $\alpha$-WP$_2$.

In Figs. 2(e)-(f) the normalized phonon frequencies for most phonon modes are plotted. The low temperature phonon softening occurs with a comparable magnitude for all modes. The total phonon anharmonicity given by the difference of the frequencies in the low and high temperature limit of the phonon frequency data differs from mode to mode and shows a larger magnitude for the higher frequency phonons. At this point we emphasize that recent photo emission experiments on $\beta$-WP$_2$ did not detect a temperature dependence of the Weyl points or other electronic features for energies close to $E_\mathrm{F}$~\cite{Razzoli-2018}. Therefore, we assume that the electronic structure does not exhibit a general tendency for an instability.


\begin{figure*}
\label{electronic-scattering}
\centering
\includegraphics[width=15cm]{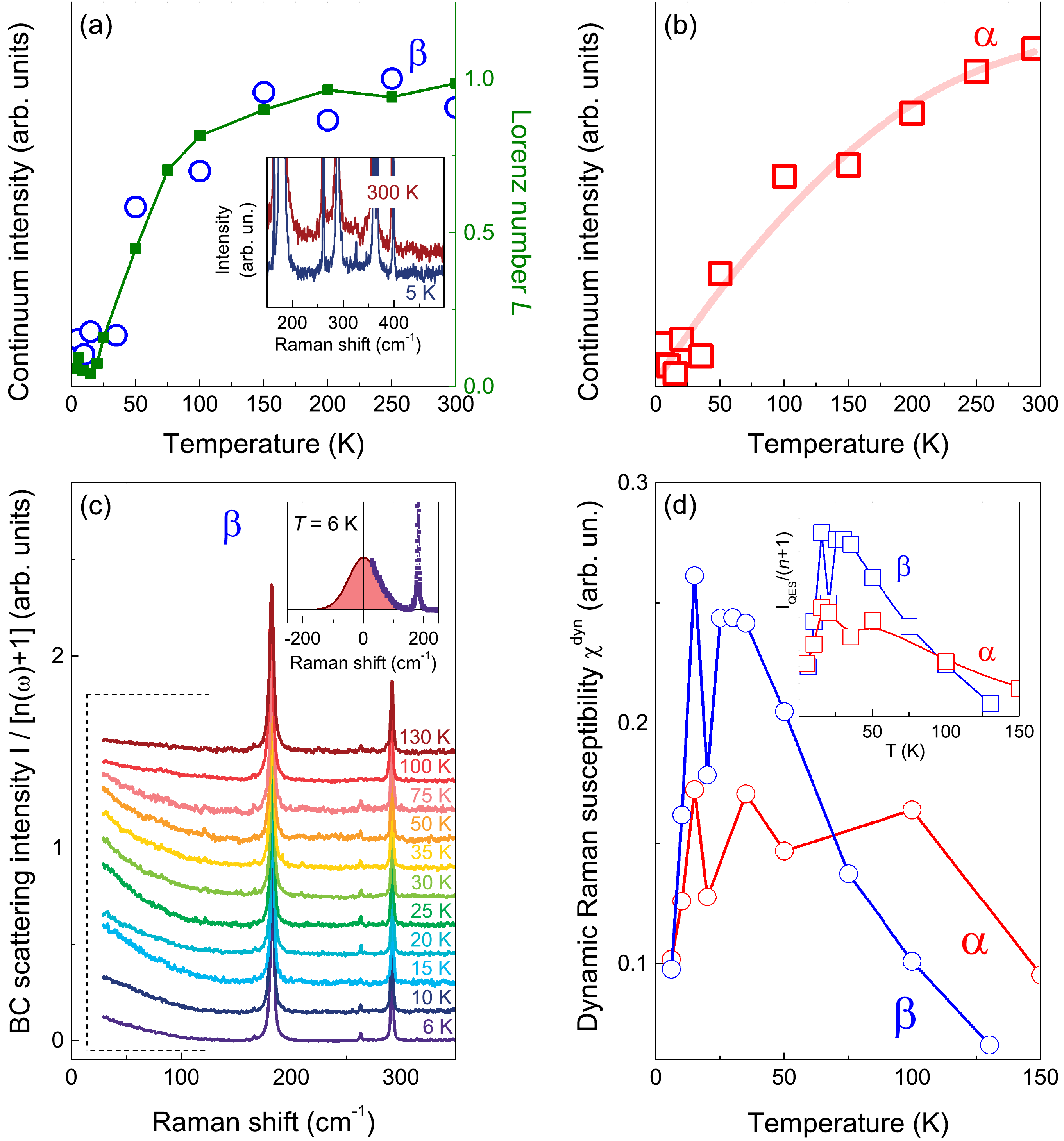}
\caption{(Color online) (a) Intensity $I(T)$ of a finite energy continuum of $\beta$-WP$_2$ as blue circles. Lorenz number $L(T)$ from thermal and electrical transport in $\beta$-WP$_2$~\cite{gooth-18} (green squares). The inset gives as-measured Raman spectra of $\beta$-WP$_2$ at 300 K (red) and 5 K (blue). (b) Intensity $I(T)$ of the continuum of $\alpha$-WP$_2$ (red squares) together with a power law fit (red line). (c) Bose corrected Raman spectra of the low-energy regime of $\beta$-WP$_2$ with quasi-elastic scattering. The inset shows a fit of the 6~K data to a Lorentzian. (d) Dynamic Raman susceptibility of both phases derived from a frequency integration of $I_{\mathrm{QES}}$ as a function of temperature. The inset shows the Bose corrected intensity, $I_{\mathrm{QES}}/(n+1)$, of both phases.}
\end{figure*}



\subsection{Finite energy and quasielastic ERS}

Fig. 3(a) and (b) follow the temperature dependence of the intensity $I(T)$ of a broad continuum that is observed at approximately 300 cm$^{-1}$ in the $\beta$ and $\alpha$ phases, respectively. An inset shows a zoom into the data of the $\beta$-phase as an example. This finite energy continuum points to multiparticle processes, most probably electronic interband scattering in the semimetal energy landscape. Phonon scattering is not taken into account as there is no evidence for a lifting of momentum conservation that could induce a continuum.

At room temperature these continua share a similar intensity in both phases and also their polarization vectors are both aligned parallel to the chain of WP coordinations, i.e. along the \textit{a} axis in the $\beta$-phase and along the \textit{b} axis in the $\alpha$-phase. In the $\beta$-phase and for temperatures below 60 K, $I(T)$ is constant and small. For higher temperatures it is much larger and develops a kind of plateau. This evolution is very close to the temperature dependence of the Lorenz number $L(T)$ of $\beta$-WP$_2$ derived from thermal and electrical transport~\cite{gooth-18}, see solid green squares. In the $\alpha$-phase $I(T)$ shows a continuous temperature dependence and can be fitted to a power law. Here, no detailed thermal transport data exists so far for a comparison.

In Fig. 3(c) and (d) quasielastic scattering (QES, $E \approx$~0) observed in the two phases of WP$_2$ is depicted, see Fig. 3(c) with Bose corrected data from the $\beta$-phase and an inset with a Lorentzian fit. The intensity of QES is larger than that of the finite energy continuum. It has the same selection rules as the continuum. However, for the $\alpha$-phase QES is also observable with polarizations perpendicular to the chain direction. The inset of Fig. 3(d) gives the Bose corrected intensity, $I_{\mathrm{QES}}/(n+1)$, of QES with the Bose factor $n(\omega,T)$. For $\beta$-WP$_2$ we find a pronounced and asymmetric peak from 10 -- 50~K, with a maximum at $T_{\mathrm{max}} = 15$ K. At lowest temperatures the data drops to a minimum comparable to the room temperature value. At high temperatures there exists a linear decrease with temperature. For $\alpha$-WP$_2$ $I_{\mathrm{QES}}/(n+1)$ is less pronounced with a more shallow maximum and a flatter high temperature tail.

ERS in simple metals is related to electronic density fluctuations due to intraband excitations and hardly observable due to its small energy, $E \approx$~0, and screened intensity~\cite{Klein-1983}. Only in bad metals with larger scattering rates it may form a finite energy maximum with the peak position related to the scattering rate. Finally, in the limit of a marginal Fermi liquid with diffusive dynamics~\cite{Varma-2017} ERS can be observed over an extended continuum of energy. We refer here to high temperature superconductors based on cuprates~\cite{Devereaux-review} and ferropnictides~\cite{Hackl-review}. For the latter nesting is also an important ingredient. QES and low energy maxima are also observed in topological insulators \cite{gnezdilov-11,gnezdilov-14}, the Dirac semimetal Cd$_3$As$_2$~\cite{sharafeev-17}, as well as in single and bilayer graphene \cite{Ponosov-2015,Riccardi-2016,Riccardi-2019}. The intensity $I(T)$ of ERS in Cd$_3$As$_2$ is similar to the data shown in Fig.~3. For a microscopic approach to this scattering in topological semimetals we refer to the Appendix detailing the evaluation of scattering Hamiltonians.

A Bose correction with $I(\omega)$= [1 + $n(\omega,T)$] $\chi''(\omega)$ leads to the Raman susceptibility $\chi''(\omega)$ and the Raman conductivity $\chi''(\omega)/\omega$. In an ansatz analogous to the Kramers-Kronig relation a dynamic Raman susceptibility is defined, $\chi^{\mathrm{dyn}}=\lim_{\omega \rightarrow 0} \chi(k=0,\omega)\equiv\frac{2}{\pi}\int^{\infty}_{0}\frac{\chi''(\omega)}{\omega}d\omega$. The latter quantity successfully describes dynamic electronic properties, e.g. inter-/intraband fluctuations in Fe-based superconductors that are complex due to the multi-sheeted Fermi surface and pronounced nesting~\cite{Gallais}.

For the $\beta$-phase $\chi^{\mathrm{dyn}}$ shows a broad, asymmetric maximum at $T_{\mathrm{max}} = 25$~K with a sharp drop at low temperatures, see Fig.~3(d), blue curve. For the $\alpha$-phase the maximum is less obvious and shifted to higher temperatures (red curve). We find the enhanced $\chi^{\mathrm{dyn}}$ of the $\beta$- with respect to the $\alpha$-phase most significant as it has to be related to enhanced electronic fluctuations. This effect is also obvious in the Bose corrected data. In addition to this data analysis we checked for characteristic effects based on energy density fluctuations, e.g., due to some order suppressed by low dimensionality~\cite{Lemmens-review} and to scaling of the data known from precursors of a phase transition~\cite{Glamazda-2016}. The available data does not support such approaches. Therefore we do not discuss them here anymore.


\section{Discussion}


The semimetal WP$_2$ with the two phases, $\alpha$-WP$_2$ and $\beta$-WP$_2$, offers the unique opportunity to evaluate the effects of topological invariants on electronic fluctuations and phonons. The most prominent shared properties are given by anomalous deviations of the optical phonons from anharmonicity, i.e., the low temperature softening and the intermediate temperature broadening in the interval 50 -- 350~K. These anomalies are also evident in the phonon intensity and they are not restricted to a particular mode. This means that they do not evidence a lattice instability of a particular form. Instead, for both phases they reflect the consequence of low dimensionality given by the chain of tungsten ions and the very anisotropic Fermi surfaces with evidence for nesting~\cite{du-18,gooth-18}. A conventional structural instability, however, is prohibited by the low site symmetry that does not support releasing degeneracies. This is consistent with the absence of temperature dependent features of the Fermi surface~\cite{Razzoli-2018}.

More subtle differences exist in an enhanced magnitude of these effects for the non-topological $\alpha$-phase. This means that the non-topological phase has a larger electron-phonon coupling and accordingly, that this is not essential for transport anomalies of the topological phase. This conclusion is diametrical to a theoretical consideration of phonon scattering processes in $\beta$-WP$_2$ \cite{Su-2019}. On the other side, the observed anomalies do not contradict the present knowledge of topologically induced phonon anomalies~\cite{garate-13, saha-13}. A general approach for semimetals shows that the respective matrix elements have a strong momentum dependence and differentiate inter- vs intraband transitions [see inset in Fig. 1(a)]. Intraband fluctuations with small $k$ are related to weaker phonon anomalies, while interband fluctuations couple rather efficiently to phonons at large $k$. We attribute the observed finite temperature effects in WP$_2$ to deviations from conventional anharmonicity involving phonon decay processes. These processes involve phonons with large momenta up to the zone boundary. The Bose factor $n(\omega,T)$ that describes their thermal occupation leads to a low temperature onset of additional linewidth for $T \gtrapprox$ 50~K. The high frequency cutoff of the acoustic phonon branches at the zone boundary is compatible with the upper temperature limit of the anomalies. Together, this leads to a finite temperature interval of enhanced linewidth. Such well-defined intervals of phonon anomalies~\cite{Yan-2018} together with the corresponding ERS~\cite{Riccardi-2016, Riccardi-2019} have already been observed in electrically tuned bilayer graphene.


In the following we will discuss ERS as the second shared property of the WP$_2$ phases. We tentatively attribute the finite energy continuum to interband and the QES to intraband scattering, respectively. This is based on their different typical energy scales. The similarity of $I(T)$ of QES in the $\beta$-phase to its Lorenz number $L(T)$ therefore results from intraband fluctuations being a dominant scattering process at low temperatures (in the hydrodynamic transport regime). Its evolution consists of a gradual narrowing of the quasiparticle peak with decreasing temperatures leading to a reduced spectral weight in the energy window of the Raman scattering experiments.

It is interesting to note that the topological $\beta$-phase shows a more pronounced dependence $I(T)$ compared to the $\alpha$-phase, despite larger phonon anomalies in the latter. On the other side, one would expect a more pronounced temperature dependence as the low temperature regime is characterized by hydrodynamic charge transport. At this point we refer to the drastic enhancement and drop of $\chi^{\mathrm{dyn}}$ above $T^*$ and preceding the temperature coherent charge transport. The underlying finite energy RS continuum is a rather rare observation for a metal and attributed to diffusive dynamics or marginal quasiparticles. The latter are attributed to low-dimensionality and nesting of the Fermi surface. The drop in $I(T)$ of the finite energy continuum corresponds to a crossover from a marginal Fermi liquid to a coherent regime that has frequently been discussed in strongly correlated systems~\cite{Varma-2017}. The observed hydrodynamic transport is a result of the coherence probed by interband fluctuations in RS.

The implications of low-dimensionality and nesting of the Fermi surface are furthermore highlighted by the selection rules of the electronic Raman scattering. Its intensity is observed with incident and scattered photons polarized parallel to the chain direction. In contrast, a microscopic evaluation of scattering rates and intensities based on Green's functions for isotropic electronic bands leads to a dominance of crossed polarizations, see Supplement. In this sense the two semimetal phases of WP$_2$ are rather special. 

Summarizing, we probe a dichotomy of interband and intraband scattering processes in WP$_2$ on different energy scales. The temperature dependencies of the respective Raman scattering intensities show gradual as well as more abrupt changes with temperature that herald the regime of hydrodynamic transport. Electron-phonon coupling is significant for both phases. However, it dominates in the non-topological phase of WP$_2$.

\begin{acknowledgments}
We acknowledge important discussions with Yann Gallais, Doohee Cho, and Yurii Pashkevich. Work done by D.W. and P.L. was supported by ``Nieders\"{a}chsisches Vorab'' through ``Quantum- and Nano-Metrology (QUANOMET), NL-4'', and by Deutsche Forschungsgemeinschaft (DFG, German Research Foundation) DFG LE967/16-1 and EXC-2123 QuantumFrontiers – 390837967. Samples were provided for this work under the ERC Advanced Grant No. 742068 ’TOPMAT’.
\end{acknowledgments}





\section{Supplement}

\subsection{Phonon analysis}
For orthorhombic $\beta$-WP$_2$ in $Cmc_21$ with one W- and two P-ions located on $4a$ Wyckoff positions a factor-group analysis leads to $5 \cdot A_1 + 3 \cdot A_2 + 2 \cdot B_1 + 5 \cdot B_2 = 15$ Raman-active optical phonon modes. The experimentally determined phonon frequencies and linewidths are detailed in Table I.

Monoclinic $\alpha$-WP$_2$ in $C12/m1$ with one W- and two P-ions each located on $4i$ Wyckoff positions~\cite{ruhl} yields $6 \cdot A_g + 3 \cdot B_g = 9$ Raman-active optical phonon modes. Details on the phonon assignment, energies and linewidths can be found in Table II.

\begin{table*}
	\caption{\label{tab:table1}Optical phonon modes of the topological orthorhombic $\beta$-WP$_2$ in $au$ polarization. The observed linewidths in FWHM are determined in the low temperature limit and lead to a mean linewidth of $\Delta\omega_{\mathrm{mean}}$=2.6~cm$^{-1}$.}
	\begin{ruledtabular} \begin{tabular}{ c|c|c|c|c }
			$\omega_{\mathrm{exp}}$ (cm$^{-1}$)&Assignment&Lineshape&Linewidth (cm$^{-1}$)\\ \hline
			164&$A_2$&Lorentz&            1.9\\
			174&$B_2$&Lorentz&            3.5\\
			180&$A_1$&Lorentz&            3.0\\
			261&$B_1 / A_2$&Lorentz&	1.9  \\
			290&$A_1$&Lorentz&            2.4\\
			297&$B_2 / A_2$&Lorentz&	2.0  \\
			325&$B_1$&Lorentz&            2.4\\
			360&$A_1$&Lorentz&            2.7\\
			367&$B_2$&Lorentz&            2.5\\
			399&$A_1$&Lorentz&            2.2\\
			440&$B_2$&Lorentz&            3.8\\
			520&$A_1$&Lorentz&            2.2\\
			550&$B_2$&Lorentz&            3.0\\
\end{tabular} \end{ruledtabular} \end{table*}

\begin{table*}
	\caption{\label{tab:table2}Optical phonon modes of the topologically trivial monoclinic $\alpha$-WP$_2$ in $bu$ polarization. The observed linewidths in FWHM are determined in the low temperature limit and lead to a mean linewidth of $\Delta\omega_{\mathrm{mean}}$=3.4~cm$^{-1}$. The observed linewidth in the low temperature limit is given as FWHM.}
	\begin{ruledtabular} \begin{tabular}{ c|c|c|c }
			$\omega_{\mathrm{exp}}$ (cm$^{-1}$)&Assignment&Lineshape&Linewidth (cm$^{-1}$)\\ \hline
			145&$A_g / B_g$&Fano&         2.0\\
			225&$A_g$&Fano&               2.4\\
			279&$A_g / B_g$&Lorentz&      4.3\\
			302&$B_g$&Fano&               3.8\\
			388&$A_g$&Lorentz&            4.0\\
			418&$A_g$&Lorentz&            2.5\\
			508&$A_g$&Lorentz&            4.5\\
\end{tabular} \end{ruledtabular} \end{table*}

\subsection{ERS in Topological Semimetals}

Based on transport experiments and theoretical considerations hydrodynamic charge dynamics exist in the topological phase WP$_2$. To account for elastic scattering of intraband electron we take into consideration particles and holes in the Weyl semimetals:

\subsubsection{Scattering rates 1/$\tau$}

In Weyl semimetals the particle and hole contribution for small $ q$ is given by the chiral Green's function  $G_{L}(\vec{k},i\nu_{n})$ (for the left chirality) and $G_{R }(\vec{k},i\nu_{n})$ (for the right chirality):

$G_{L}(\vec{k}+\vec{q},i\nu_{n})=\frac{1}{i\nu_{n}+\mu -\epsilon(\vec{k}+\vec{q})}+\frac{1}{i\nu_{n}+\mu +\epsilon(\vec{k}+\vec{q})}\approx\frac{1}{i\nu_{n}-\epsilon(\vec{k})+\mu}+\frac{1}{i\nu_{n}+\epsilon(\vec{k})+\mu}$

The dispersion for the left chirality, $\epsilon(\vec{k})=\sqrt{ v^2(k^2+(k_{3}-Q_{3})^2)}$,$Q_{3}$. $Q_{3}$ is the momentum of the node. In the experiment we have several nodes.

\begin{eqnarray}
&&\sum (\vec{k},i\nu_{n})=-\frac{1}{\beta}\int\frac{d^3q}{(2\pi)^3}\sum_{n}\Big(G_{L}(\vec{k}+\vec{q},i\nu_{n})
\Big)D(\Omega_{q},i\nu_{n}-i\omega_{n})\nonumber\\&&\approx
\int\frac{d^3q}{(2\pi)^3}\frac{1}{2\pi i}\int_{c}G_{L}(\vec{k},z)
D(\Omega_{q},i\nu_{n}-z) f(z)dz\nonumber\\&&
=\int\frac{d^3q}{(2\pi)^3}\frac{1}{2\pi i}\int_{c}f(z)\Big(\frac{1}{z-\epsilon(\vec{k})+\mu}+\frac{1}{z+\epsilon(\vec{k})+\mu}\Big)\frac{2\Omega(q)}{(-i\nu_{n}-i\omega_{n})^2-\Omega(q)^2},\nonumber\\&&
with ~f(z)=\frac{1}{e^{\beta z}+1} ~and~ 
D(\Omega_{q},i\nu_{n}-i\omega_{n})=\frac{2\Omega_{q}}{(i\nu_{n}-i\omega_{n})^2-\Omega^2_{q}}.\nonumber\\&&
\end{eqnarray}

We compute the contribution from the poles, perform the analytic continuation, and find for the frequency $\omega$ at finite temperature:

\begin{eqnarray}
&&\frac{1}{\tau}=\int\frac{d^3q}{(2\pi)^3}\nonumber\\&&\Big[\Big(f(-\epsilon(\vec{k})+\mu)\frac{2\Omega_{q}}{(\omega-\epsilon(\vec{k})+\mu)^2-
	 \Omega^2_{q}+i\eta}\Big)+\Big(f(-\epsilon(\vec{k})-\mu)\frac{2\Omega_{q}}{(\omega-\epsilon(\vec{k})+\mu)^2-
	\Omega^2_{q}+i\eta}\Big)\nonumber\\&&
+	 \Big(\frac{1}{1-e^{\beta\Omega_{q}}}\Big)\Big(\frac{1}{\omega-\epsilon(\vec{k})+\mu+\Omega_{q}+i\eta}+ \frac{1}{\omega-\epsilon(\vec{k})+\mu+\Omega_{q}+i\eta}\Big)\nonumber\\&&\Big(\frac{1}{1-e^{-\beta\Omega_{q}}}\Big)\Big(\frac{1}{\omega-\epsilon(\vec{k})+\mu- \Omega_{q}+i\eta}+\frac{1}{\omega+\epsilon(\vec{k})+\mu- \Omega_{q}+i\eta}\Big)\Big].\nonumber\\&&
\end{eqnarray}


Ignoring the $Ferm$i-$Dirac$ contribution, we obtain at low temperature:
\begin{eqnarray}
	&&\frac{1}{\tau}=Im.\sum (\vec{k},\omega)=\frac 
	{g^2_{e_ph}}{2(2\pi)^2}\int\,d^3q\Big[ 
	\Big(\frac{1}{1-e^{\beta\Omega_{q}}}\Big) \delta(\omega 
	-\epsilon(\vec{k})+\mu+\Omega_{q})\nonumber\\&& 
	-\Big(\frac{1}{1-e^{\beta\Omega_{q}}}+1\Big)\delta(\omega 
	-\epsilon(\vec{k})+\mu-\Omega_{q})\Big]. \nonumber\\&&
\end{eqnarray}

For acoustic phonons, $\Omega_{q}\approx C|q|$, the scattering rate is: 

$\frac{1}{\tau}=\frac {g^2_{e_ph}}{C^32(2\pi)^2}\Big[(\omega-\epsilon(\vec{k})+\mu)^2(\frac{1}{1-e^{\hbar v \beta(\omega-\epsilon(\vec{k})+\mu)}})-(\omega-\epsilon(\vec{k})+\mu)^2(\frac{1}{1-e^{\hbar v \beta(-\omega+\epsilon(\vec{k})-\mu)}})\Big]
\propto T$


For optical phonons, $ \Omega_{q}=\Omega_{0}-\alpha q^2$, the scattering 
rate is:

\begin{eqnarray}
	&&
	\frac{1}{\tau}=\frac 
	{g^2_{e_ph}}{45(2\pi)^2}\Big[\sqrt{(-\omega+\epsilon(\vec{k})-\mu+\Omega_{0})}\Big(\frac{1}{1-e^{\hbar 
			v \beta(-\omega+\epsilon(\vec{k})-\mu)}}\Big)\nonumber\\&& 
	-\sqrt{(\omega-\epsilon(\vec{k})+\mu-\Omega_{0})}(\frac{1}{1-e^{\hbar v 
			\beta(\omega-\epsilon(\vec{k})+\mu)}})\Big].\nonumber\\&&
\end{eqnarray}

\subsubsection{Raman Scattering in the topological phase}

The excited bands are of Left $L$ and Right $R$ chirality. Phonons couple to intraband excitations and we have $L^{\dagger}L~phonon$, $R^{\dagger}R~phonon$ and h.c. Light couples the ground state (the valence band) to the excited Left or Right chiral bands  (ground state), respectively. This leads to the $V^{\dagger}L~photon$, $V^{\dagger}R~photon$ and h.c. The resulting triangle diagrams have important implications for the selection rules of light scattering, i.e. an effect of the polarization of light in the Raman scattering experiment. We notice that significant scattering will occur when $\vec{E}_{\mathrm{in}}$ is perpendicular to $\vec{E}_{\mathrm{out}}$. For this case we have the unique chance to probe the winding number $J=1,2$ introduced by J. P. Carbotte \cite{Mukh-2018} for a topological semimetal using namely this experiment.

The scattering involves separately $Left$, $Right$ excited chiral bands and the valence band ground state given by the Green's function $g_{valence}$. The Green's functions of the chiral electrons are $G_{Left}$ and $G_{Right}$.

\begin{eqnarray}
&&S_{L}^{(3)}=\int\,d^{3}x_{1}\int\,dt_{1}\int\,d^{3}x_{2}\int\,dt_{2}\int\,d^{3}x_{3}\int\,dt_{3}\nonumber\\&& \times Tr\Big[G_{Left}(\vec{x_{2}},t_{2};\vec{x_{1}},t_{1})\varphi(\vec{x_{2}},t_{2})G_{Left}(\vec{x_{3}},t_{3};\vec{x_{2}},t_{2})\sigma_{1}
\cdot\vec{A}_{1}(\vec{x_{3}},t_{3})g_{valence}(\vec{x_{1}},t_{1};\vec{x_{3}},t_{3})\sigma_{2}
\cdot\vec{A}_{2}(\vec{x_{1}},t_{1})\Big] \nonumber\\&&
\end{eqnarray}
A similar contribution comes from the other chiral band giving $S_{R}^{(3)}$. The Raman intensity is then given by $I=|S_{L}^{(3)}+S_{R}^{(3)}|^2$.

A comparison with our experimental data shows that electronic Raman scattering in WP$_2$ mainly evolves in polarization parallel to the chain-like crystallographic coordination. We attribute this to the strongly anisotropic band structure that has presently not been taken into account.

\end{document}